\documentclass[a4paper,manyauthors,nocleardouble,COMPASS]{cernphprep}
\usepackage{etex}
\usepackage{amsmath,amssymb,fleqn,url,color,graphics,graphicx}
\usepackage{enumerate}
\usepackage{bigstrut}
\usepackage{soul,xcolor}
\usepackage{cite}
\usepackage{lineno, xcolor}
\usepackage{hyperref}

\hypersetup{
    colorlinks,
    citecolor=blue,
    filecolor=blue,
    linkcolor=blue,
    urlcolor=blue
}

\newcommand{\gvc}{GeV/$c$ }
\newcommand{\gvcs}{(GeV/$c$)$^2$}
\newcommand{\gvcw}{GeV/$c^2$}
\newcommand{\GeV}{\mathrm{GeV}}

\begin{document}
\begin{titlepage}

\PHnumber{2016--250}
\PHdate{
  \vspace*{-1.1\baselineskip}
  \begin{tabular}[c]{r@{}}
  23 September 2016\\
  rev.\ Mar.\ 8 2017
  \end{tabular}
}

\title{Sivers asymmetry extracted in SIDIS at the hard scales of the Drell-Yan process at COMPASS}

\Collaboration{The COMPASS Collaboration}
\ShortAuthor{The COMPASS Collaboration}

\begin{abstract}
Eight proton transverse-spin azimuthal asymmetries are extracted in four regions of the photon virtuality $Q^2$ from the COMPASS 2010 semi-inclusive hadron measurements in deep inelastic muon-nucleon scattering. These $Q^2$ regions correspond to the four regions of the di-muon mass $\sqrt{Q^2}$ used in the ongoing analyses of the COMPASS Drell-Yan measurements, which allows for a future direct comparison of the nucleon transverse-momentum-dependent parton distribution functions extracted from these two alternative measurements. In addition, for the azimuthal asymmetries induced by the Sivers transverse-momentum-dependent parton distribution function various two-dimensional kinematic dependences are presented. The integrated Sivers asymmetries are found to be positive with an accuracy that appears to be sufficient to test the sign change of the Sivers function predicted by Quantum Chromodynamics.
\end{abstract}

\vspace*{60pt}

\vfill
\Submitted{(to be submitted to Phys.\ Lett.\ B)}

\end{titlepage}

{\pagestyle{empty} 
%
%
\section*{The COMPASS Collaboration}
\label{app:collab}
\renewcommand\labelenumi{\textsuperscript{\theenumi}~}
\renewcommand\theenumi{\arabic{enumi}}
\begin{flushleft}
C.~Adolph\Irefn{erlangen},
M.~Aghasyan\Irefn{triest_i},
R.~Akhunzyanov\Irefn{dubna}, 
M.G.~Alexeev\Irefn{turin_u},
G.D.~Alexeev\Irefn{dubna}, 
A.~Amoroso\Irefnn{turin_u}{turin_i},
V.~Andrieux\Irefnn{illinois}{saclay},
N.V.~Anfimov\Irefn{dubna}, 
V.~Anosov\Irefn{dubna}, 
K.~Augsten\Irefnn{dubna}{praguectu}
% ...
\end{flushleft}
%
%
\begin{Authlist}
\item \Idef{aveiro}{University of Aveiro, Dept.\  of Physics, 3810-193 Aveiro, Portugal}
\item \Idef{bochum}{Universit\"at Bochum, Institut f\"ur Experimentalphysik, 44780 Bochum, Germany\Arefs{l}\Arefs{s}}
\item \Idef{bonniskp}{Universit\"at Bonn, Helmholtz-Institut f\"ur  Strahlen- und Kernphysik, 53115 Bonn, Germany\Arefs{l}}
\item \Idef{bonnpi}{Universit\"at Bonn, Physikalisches Institut, 53115 Bonn, Germany\Arefs{l}}
\item \Idef{brno}{Institute of Scientific Instruments, AS CR, 61264 Brno, Czech Republic\Arefs{m}}
\item \Idef{calcutta}{Matrivani Institute of Experimental Research \& Education, Calcutta-700 030, India\Arefs{n}}
\item \Idef{dubna}{Joint Institute for Nuclear Research, 141980 Dubna, Moscow region, Russia\Arefs{o}}
\item \Idef{erlangen}{Universit\"at Erlangen--N\"urnberg, Physikalisches Institut, 91054 Erlangen, Germany\Arefs{l}}
\item \Idef{freiburg}{Universit\"at Freiburg, Physikalisches Institut, 79104 Freiburg, Germany\Arefs{l}\Arefs{s}}
\item \Idef{cern}{CERN, 1211 Geneva 23, Switzerland}
\item \Idef{liberec}{Technical University in Liberec, 46117 Liberec, Czech Republic\Arefs{m}}
\item \Idef{lisbon}{LIP, 1000-149 Lisbon, Portugal\Arefs{p}}
\item \Idef{mainz}{Universit\"at Mainz, Institut f\"ur Kernphysik, 55099 Mainz, Germany\Arefs{l}}
\item \Idef{miyazaki}{University of Miyazaki, Miyazaki 889-2192, Japan\Arefs{q}}
\item \Idef{moscowlpi}{Lebedev Physical Institute, 119991 Moscow, Russia}
\item \Idef{munichtu}{Technische Universit\"at M\"unchen, Physik Dept., 85748 Garching, Germany\Arefs{l}\Arefs{r}}
\item \Idef{nagoya}{Nagoya University, 464 Nagoya, Japan\Arefs{q}}
\item \Idef{praguecu}{Charles University in Prague, Faculty of Mathematics and Physics, 18000 Prague, Czech Republic\Arefs{m}}
\item \Idef{praguectu}{Czech Technical University in Prague, 16636 Prague, Czech Republic\Arefs{m}}
\item \Idef{protvino}{State Scientific Center Institute for High Energy Physics of National Research Center `Kurchatov Institute', 142281 Protvino, Russia}
\item \Idef{saclay}{IRFU, CEA, Universit\'e Paris-Saclay, 91191 Gif-sur-Yvette, France\Arefs{s}}
\item \Idef{taipei}{Academia Sinica, Institute of Physics, Taipei 11529, Taiwan}
\item \Idef{telaviv}{Tel Aviv University, School of Physics and Astronomy, 69978 Tel Aviv, Israel\Arefs{t}}
\item \Idef{triest_u}{University of Trieste, Dept.\  of Physics, 34127 Trieste, Italy}
\item \Idef{triest_i}{Trieste Section of INFN, 34127 Trieste, Italy}
\item \Idef{triest_ictp}{Abdus Salam ICTP, 34151 Trieste, Italy}
\item \Idef{turin_u}{University of Turin, Dept.\  of Physics, 10125 Turin, Italy}
\item \Idef{turin_i}{Torino Section of INFN, 10125 Turin, Italy}
\item \Idef{illinois}{University of Illinois at Urbana-Champaign, Dept.\  of Physics, Urbana, IL 61801-3080, USA}
\item \Idef{warsaw}{National Centre for Nuclear Research, 00-681 Warsaw, Poland\Arefs{u} }
\item \Idef{warsawu}{University of Warsaw, Faculty of Physics, 02-093 Warsaw, Poland\Arefs{u} }
\item \Idef{warsawtu}{Warsaw University of Technology, Institute of Radioelectronics, 00-665 Warsaw, Poland\Arefs{u} }
\item \Idef{yamagata}{Yamagata University, Yamagata 992-8510, Japan\Arefs{q} }
\end{Authlist}
%
%
\renewcommand\theenumi{\alph{enumi}}
\begin{Authlist}
\item [{\makebox[2mm][l]{\textsuperscript{\#}}}] Corresponding authors
\item [{\makebox[2mm][l]{\textsuperscript{*}}}] Deceased
\item \Adef{a}{Also at Instituto Superior T\'ecnico, Universidade de Lisboa, Lisbon, Portugal}
\item \Adef{b}{Also at Dept.\ of Physics, Pusan National University, Busan 609-735, Republic of Korea and at Physics Dept., Brookhaven National Laboratory, Upton, NY 11973, USA}
\item \Adef{r}{Supported by the DFG cluster of excellence `Origin and Structure of the Universe' (www.universe-cluster.de)}
\item \Adef{d}{Also at Chubu University, Kasugai, Aichi 487-8501, Japan\Arefs{q}}
\item \Adef{x}{Also at Dept.\  of Physics, National Central University, 300 Jhongda Road, Jhongli 32001, Taiwan}
\item \Adef{e}{Also at KEK, 1-1 Oho, Tsukuba, Ibaraki 305-0801, Japan}
\item \Adef{g}{Also at Moscow Institute of Physics and Technology, Moscow Region, 141700, Russia}
\item \Adef{v}{Supported by Presidential grant NSh--999.2014.2}
\item \Adef{turin_p}{Also at University of Eastern Piedmont, 15100 Alessandria, Italy}
\item \Adef{h}{Present address: RWTH Aachen University, III.\ Physikalisches Institut, 52056 Aachen, Germany}
\item \Adef{y}{Also at Dept.\  of Physics, National Kaohsiung Normal University, Kaohsiung County 824, Taiwan}
\item \Adef{i}{Present address: Uppsala University, Box 516, 75120 Uppsala, Sweden}
\item \Adef{c}{Supported by the DFG Research Training Group Programmes 1102 and 2044} 
%
%
\item \Adef{l}{Supported by the German Bundesministerium f\"ur Bildung und Forschung}
\item \Adef{s}{Supported by EU FP7 (HadronPhysics3, Grant Agreement number 283286)}
\item \Adef{m}{Supported by Czech Republic MEYS Grant LG13031}
\item \Adef{n}{Supported by SAIL (CSR), Govt.\ of India}
\item \Adef{o}{Supported by CERN-RFBR Grant 12-02-91500}
\item \Adef{p}{\raggedright Supported by the Portuguese FCT - Funda\c{c}\~{a}o para a Ci\^{e}ncia e Tecnologia, COMPETE and QREN,
 Grants CERN/FP 109323/2009, 116376/2010, 123600/2011 and CERN/FIS-NUC/0017/2015}
\item \Adef{q}{Supported by the MEXT and the JSPS under the Grants No.18002006, No.20540299 and No.18540281; Daiko Foundation and Yamada Foundation}
\item \Adef{t}{Supported by the Israel Academy of Sciences and Humanities}
\item \Adef{u}{Supported by the Polish NCN Grant 2015/18/M/ST2/00550}
\end{Authlist}
}
\newpage
\setcounter{page}{1}
\parindent=0em

%
%
%
\section{Introduction}
\label{sec:intro}
Parton distribution functions (PDFs) play a very important role in the theoretical description of high energy reactions. Recent decades were marked by enormous progress in both theoretical and experimental studies of spin-(in)dependent and transverse-momentum-dependent (TMD) nucleon PDFs. The latter provide a three-dimensional picture of a fast moving nucleon in momentum space, for recent reviews see Refs.~\cite{Peng:2016zwn,Perdekamp:2015vwa,Peng:2014hta,Aidala:2012mv, Barone:2010zz}. The TMD factorisation was proven to hold~\cite{Collins:2011zzd} for the cross sections of semi-inclusive measurements of hadron production in deep-inelastic lepton-nucleon scattering, $\ell \,N \rightarrow \ell^\prime \,h \, X$ (hereafter referred to as SIDIS) and of lepton-pair production in the Drell-Yan process, $h \, N \rightarrow \ell\,\bar{\ell}\, X$ (hereafter referred to as DY).
This allows comparative studies of the same nucleon TMD PDFs and their dependence on the hard scale $Q$ via TMD evolution. Here, $Q^2$ is the photon virtuality in SIDIS and $Q=\sqrt{Q^2}$ is the di-muon mass in DY.

The spin and quark-transverse-momentum structure of the nucleon is described by TMD PDFs. Among them, an important role is played by the ``twist-2'' Sivers function $f_{1T}^\perp$~\cite{Sivers:1989cc} that describes the left-right asymmetry in the distribution of partons in the nucleon with respect to the plane spanned by the directions of momentum and spin of the nucleon. A peculiar feature of the Sivers  TMD PDF predicted in Refs.~\cite{Collins:2002kn, Brodsky:2002cx, Brodsky:2002rv} is that it contributes with opposite sign to SIDIS and DY, which is considered to be an essential prediction of by Quantum Chromodynamics (QCD).
Since the contribution of the Sivers TMD PDF as a ``twist-2'' object is not suppressed at high $Q^2$, measurements of the Sivers effect at largely different hard scales can be directly compared. This opens the possibility to conclude which of the existing $Q^2$-evolution schemes describes the data best.

The Sivers effect was studied in SIDIS using transversely polarised targets at HERMES~\cite{Airapetian:2009ae}, COMPASS~\cite{Alekseev:2008aa, Adolph:2012sp,Adolph:2014zba} and JLab Hall A~\cite{Qian:2011py} and nonzero results were obtained. The typical hard scale of these fixed-target measurements, $Q\approx $ (1 -- 5)\,\gvc, is quite different from the one explored in Drell-Yan measurements of the Sivers effect using $pp$-collisions at RHIC~\cite{Adamczyk:2015gyk} with $Q\approx 80$~\gvc and 90~\gvc.

The COMPASS experiment at CERN~\cite{Abbon:2007pq, Gautheron:2010wva} is presently the only place to explore the transverse spin structure of the nucleon by either SIDIS or DY measurements, using a similar set-up and a similar transversely polarised proton target.
This opens the unique opportunity, when comparing the Sivers TMD PDFs obtained from the two alternative experimental approaches, to test the opposite-sign prediction by QCD at practically the same hard scale, thereby minimising possible bias introduced by TMD evolution.

In 2010, SIDIS hadron data were taken at COMPASS using a longitudinally polarised muon beam of 160~\gvc momentum and a transversely polarised NH$_3$ proton target. In 2015, DY data were taken using a high-intensity $\pi^-$ beam of 190~\gvc and a similar transversely polarised target.

In order to provide useful input for future global analyses that will compare TMD PDFs obtained from SIDIS data with those obtained from DY data, COMPASS extracted all  transverse-target-polarisation-dependent azimuthal asymmetries in the SIDIS cross section (hereafter referred to as TSAs), using the same four $Q^2$-ranges as those selected for the analysis of the DY data:
\begin{enumerate}[i)]
  \item $1  ~\GeV/c<Q< 2  ~\GeV/c$: ``low mass'' range, where many background processes contribute;
  \item $2  ~\GeV/c<Q< 2.5~\GeV/c$: ``intermediate mass'' range;
  \item $2.5~\GeV/c<Q< 4  ~\GeV/c:$ ``J$/\psi$ mass range'';
  \item $4  ~\GeV/c<Q< 9  ~\GeV/c$: ``high mass'' range where background processes are strongly suppressed.
\end{enumerate}

Range iv) is particularly suited to study the predicted sign change of the Sivers TMD PDF when comparing SIDIS and DY results. First, this range best fulfils the requirement of TMD factorisation that the transverse momentum of the hadron in SIDIS or of the muon pair in DY has to be much smaller than $Q$.
Secondly, both SIDIS and DY cross sections for a proton target are dominated by the contribution of $u$-quark nucleon TMD PDFs in the valence region, where the extracted Sivers TMD PDF reaches its maximum~\cite{Anselmino:2005ea, Anselmino:2008sga}.

In this Letter, the main focus will be on the Sivers effect. The present experimental and theoretical understanding of TMD PDFs and TSAs is briefly summarised in Sec.~2. In Sec.~3, data selection and analysis are described.
In Sec.~4, results on the Sivers TSAs are given for the  first time in various two-dimensional kinematic representations. Conclusions are presented in Sec.~5.
	
\section{TMD PDFs and TSAs}
\label{sec:physics_background}
The general expression for the cross section of unpolarised-hadron production in polarised-lepton SIDIS off a transversely polarised nucleon comprises eight transverse-target-polarisation-dependent modulations in the azimuthal angle $\phi_h$ of the produced hadron and/or the azimuthal angle $\phi_S$ of the target spin vector~\cite{Kotzinian:1994dv,Bacchetta:2006tn}. These angles are defined in the target rest frame with the $\hat{\bf z}$ axis along the virtual-photon momentum and the $\hat{\bf x}$ axis along the lepton transverse momentum, where transverse is meant with respect to the $\hat{\bf z}$ axis. Five of these eight modulations are independent of the lepton polarisation.

Similarly, the cross section of pion-nucleon DY lepton-pair production off a transversely polarised nucleon also comprises five transverse-target-polarisation-dependent azimuthal modulations, when the polarisations of the produced leptons are summed over~\cite{Arnold:2008kf,Gautheron:2010wva}.

The quark Sivers functions have been extracted from HERMES~\cite{Airapetian:2009ae} and COMPASS~\cite{Alekseev:2008aa, Adolph:2012sp,Adolph:2014zba} data using both collinear~\cite{Anselmino:2005ea, Anselmino:2008sga} and TMD $Q^2$-evolution approaches~\cite{Aybat:2011ta, Anselmino:2012aa, Sun:2013hua, Echevarria:2014xaa}.
In the commonly accessible range of the Bjorken-$x$ variable, the Sivers TSA at HERMES was found to be somewhat larger compared to that measured at COMPASS. Taking into account that in this range the hard scale at COMPASS is as much as two to three times larger compared to that of HERMES, this observation may indicate the influence of TMD evolution effects. In order to test this conjecture, measuring TSAs at COMPASS in various $Q^2$ regions may yield very useful input for testing the effect of TMD evolution.

In DY lepton-pair production with a transversely polarised nucleon in the initial state, a $\sin(\Phi_S)$ asymmetry is generated by the Sivers effect. Here, $\Phi_S$ is the azimuthal angle of the nucleon polarisation in the target rest frame with the $\hat{\bf z}$ axis along the beam momentum and the $\hat{\bf x}$ axis along the direction of the transverse momentum of the produced di-muon.

Among the five lepton-polarisation-independent TSAs that appear in SIDIS and DY, three are induced by the ``twist-2'' Sivers ($f_{1T}^\perp$), transversity ($h_1$), pretzelosity ($h_{1T}^\perp$) TMD PDFs, while the other two are related to various ``twist-3'' objects~\cite{Bacchetta:2006tn,Arnold:2008kf}. Similarly, three SIDIS lepton-polarisation-dependent TSAs give access to ``twist-2'' $g_{1T}$ and different ``twist-3'' TMDs.
In contrast to the  Sivers function, transversity, pretzelosity and $g_{1T}$ TMD PDFs are predicted to be genuinely universal, \textit{i.e.} their contributions do not change sign between SIDIS and DY~\cite{Collins:2011zzd}.

Recently, the first measurement of TSAs in the cross section of $W$ and $Z$ production using single-transversely polarised proton-proton collisions at RHIC was reported by the STAR collaboration~\cite{Adamczyk:2015gyk}. Comparing the data with predictions from Ref.~\cite{Kang:2009bp} they conclude that the measured Sivers asymmetry appears to be better compatible with the sign-change scenario for the Sivers TMD PDF than with the one without sign change. Note that these predictions do not include TMD evolution effects and are based on parametrisations of Sivers and unpolarised TMD PDFs that were fitted to asymmetries measured at fixed-target energies~\cite{Anselmino:2008sga}.
Because of the largely different typical hard scales accessed by fixed-target and collider experiments, it is not excluded that TMD evolution effects may play a substantial role when comparing $W$ and $Z$ production to fixed target results. For completeness we note that together with the parametrisations of TMD PDFs at initial scale, the TMD evolution approach needs additional non-perturbative input information that cannot be calculated in pQCD. For various possible choices of this input information, different predictions exist~\cite{Anselmino:2012aa, Sun:2013hua, Echevarria:2014xaa}.

Altogether, measuring the Sivers effect at COMPASS both in SIDIS and DY at a comparable hard scale will provide the most direct way to check the pQCD prediction for a sign change of the Sivers TMD PDF.

\section{Data analysis}
\label{sec:data_analysis}
The analysis presented in this Letter is performed using COMPASS SIDIS data collected in 2010 using a 160~\gvc longitudinally polarised muon beam from the CERN SPS and a transversely polarised NH$_3$ target with proton polarisation $\langle P_T\rangle\approx 0.8$ and dilution factor $\langle f\rangle\approx 0.15$, where the latter describes the fraction of polarisable material in the target. These data were already used for the extraction of the Sivers and other TSAs, see Refs.~\cite{Adolph:2012sp,Adolph:2014zba,Adolph:2012sn,Parsamyan:2013ug}, where also details on the experimental apparatus are given. In the analysis presented here, the TSAs are extracted for the first time using two-dimensional representations in $(Q^2,x)$, $(Q^2,z)$, and $(Q^2,p_T)$ to prepare for the future direct comparison with TSA results expected from the analysis of COMPASS DY data. Here, $z$ and $p_T$ are the fraction of the virtual photon energy carried by the observed hadron and the transverse component of the hadron momentum, respectively.

From the total amount of about $4\times10^{10}$ recorded events, we accept only those that have a primary vertex inside the target volume, a reconstructed incident and a reconstructed scattered muon track, and at least one outgoing hadron track. In order to equalise the beam flux through the target, it is required that extrapolated beam trajectories cross all three target cells.
The deep-inelastic scattering (DIS) regime is ensured by selecting events with $Q^2>1$~\gvcs~and excluding the region of exclusive nucleon resonance production by constraining the invariant mass of the hadronic system to be $W > \sqrt{10}$~\gvcw\, (as also done at HERMES~\cite{Airapetian:2009ae}). The restrictions on the fraction of the initial lepton energy carried by the virtual photon, $0.1 < y < 0.9$, remove events with poorly reconstructed virtual-photon energy on the low side and events with large electromagnetic radiative corrections on the high side. After the application of these selection criteria about $16\times10^7$ DIS events are available for analysis.

While all above described requirements are imposed at the event level, two more constraints are applied on the kinematic variables of every detected charged hadron. First, $p_T>0.1$~\gvc ensures a good resolution in the azimuthal angle $\phi_h$. Secondly, the requirements $z>0.1$ or $z>0.2$ are alternatively used to select hadrons produced in the current fragmentation region.
The study of these two choices is motivated by previous COMPASS results on the Sivers effect~\cite{Adolph:2012sp}.

In the analysis presented here, we use reprocessed 2010 proton data, which include improved detector calibrations and in particular better muon reconstruction efficiency. For the same kinematic region, the resulting SIDIS yield is higher by about 9\% compared to the earlier analyses~\cite{Adolph:2012sp, Adolph:2012sn}. The two analyses give consistent results. For the present analysis, the four above defined $Q^2$-ranges are used. They contain 75\%, 11\%, 11\% and 3\% of the total statistics.

\begin{figure}[tbp]
\centering
\includegraphics[width=0.495\textwidth]{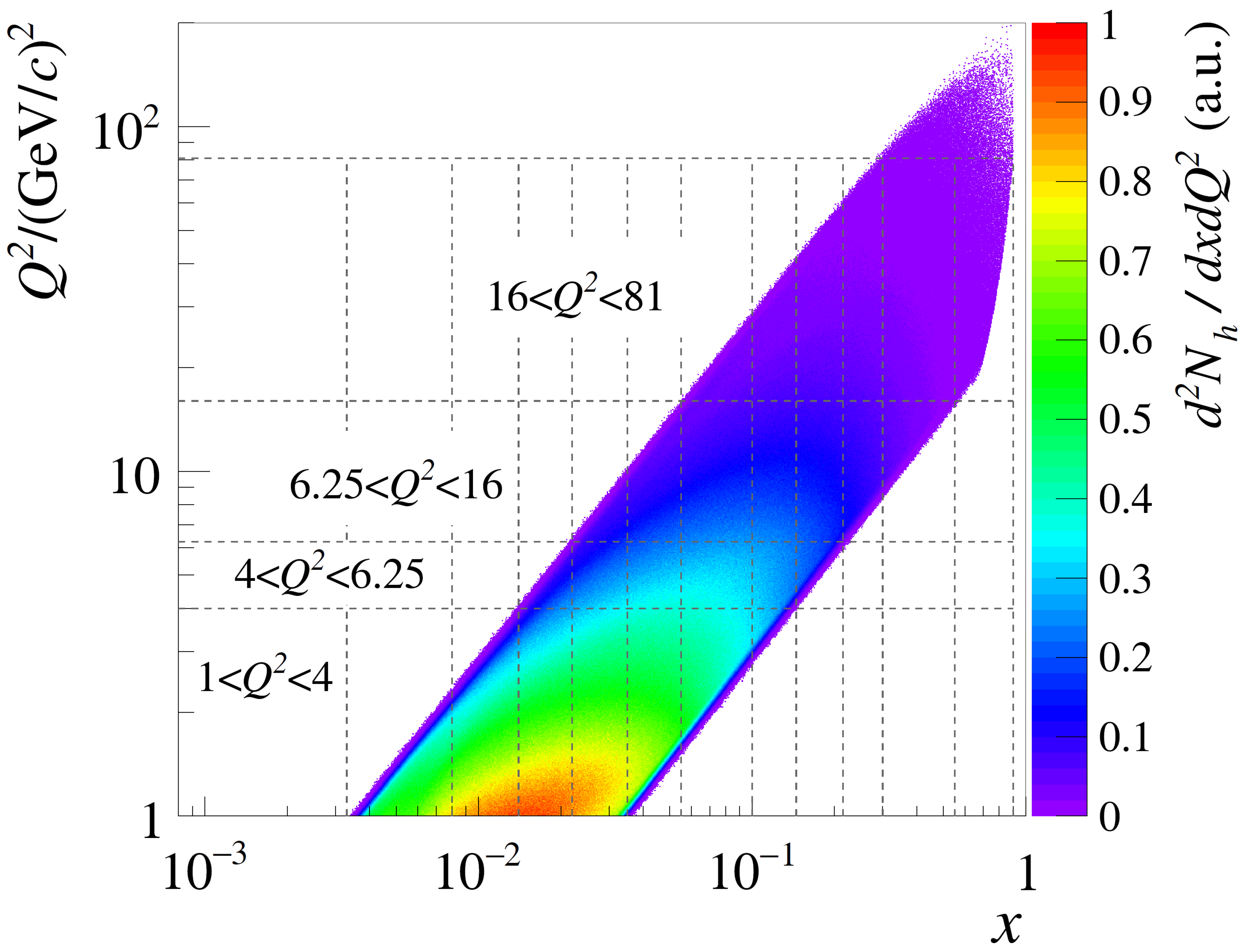}
\includegraphics[width=0.495\textwidth]{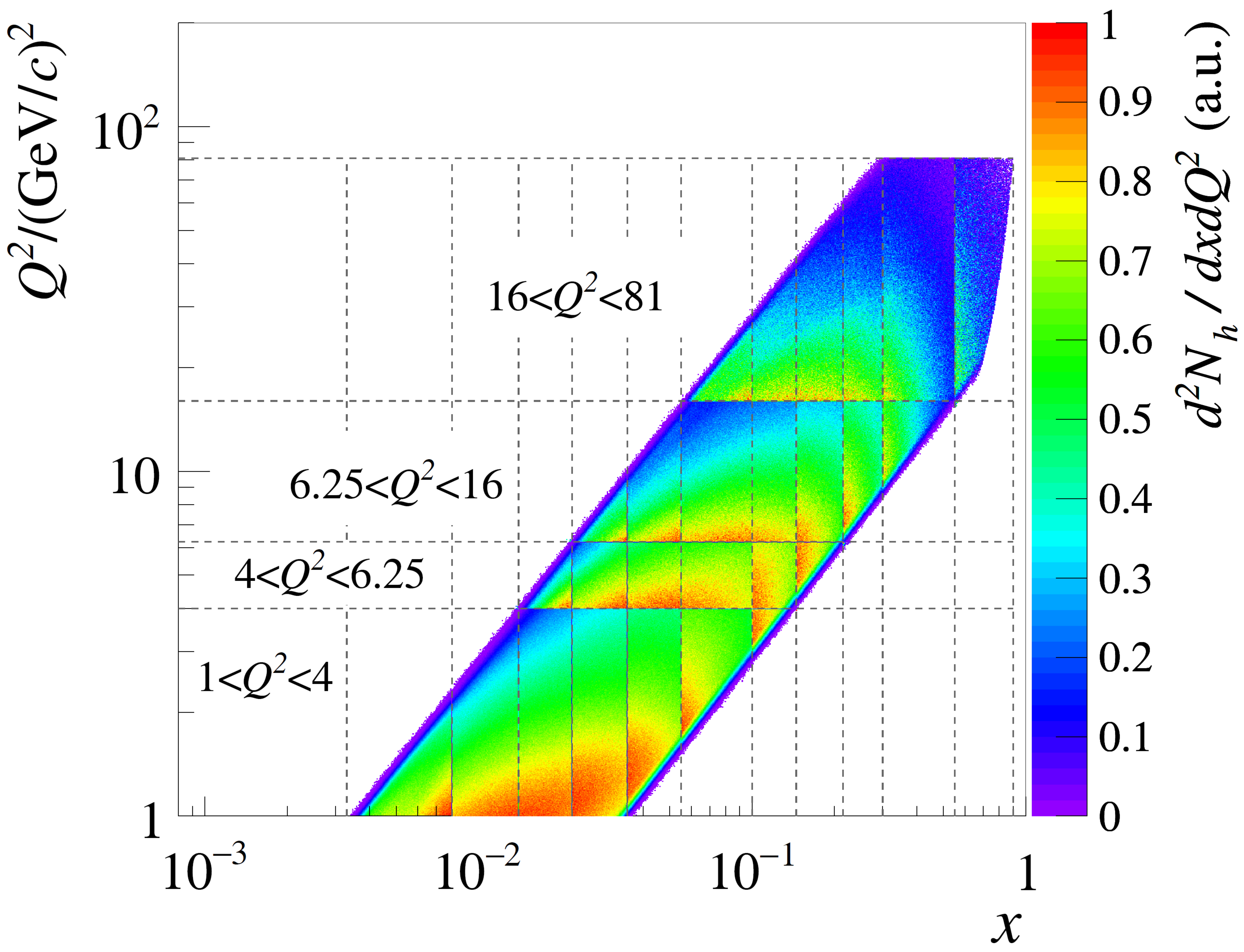}
\caption{Left panel: charged hadron SIDIS two-dimensional $(Q^2,x)$ distribution for $z>0.1$. Right panel: same distribution shown separately for each $(Q^2,x)$ cell.}
\label{fig:Q2dist}
\end{figure}

The two-dimensional $(x,Q^2)$ distribution for charged-hadron production at $z>0.1$ is shown in the left panel of Fig.~\ref{fig:Q2dist}. The distribution is normalised to have a maximum value equal to one. The right panel shows the same distribution where each $(x,Q^2)$ cell is independently normalized in the same way.

All eight TSAs that appear in the SIDIS cross section for a polarised initial lepton~\cite{Kotzinian:1994dv, Bacchetta:2006tn} are extracted simultaneously together with the corresponding correlation matrix using the extended unbinned maximum likelihood estimator as described in Ref.~\cite{Adolph:2012nw}. The lepton-polarisation-independent TSAs $A_{UT}^{w(\phi_h,\phi_S)}$ are defined as amplitudes of the azimuthal modulation $w(\phi_h,\phi_S)$ divided by the spin and azimuth-independent part of the SIDIS cross section, the effective proton polarisation ($f\cdot \langle P_T \rangle$) and the corresponding depolarisation factor. The lepton-polarisation-dependent TSAs $A_{LT}^{w(\phi_h,\phi_S)}$ are additionally divided by the beam polarisation. The subscript ($U$) $L$ denotes (in)dependence on the lepton polarisation and $T$ denotes dependence on the target transverse spin.

The TSAs are extracted separately for hadrons of positive and negative charge, where any detected hadron is counted in the analysis. With the requirement $z>0.1$, about $43\times10^6$ positive and about $34\times10^6$ negative hadrons are available for analysis, and for $z>0.2$ the numbers are approximately two times smaller.
All the results presented in this article are obtained for the range $z>0.1$. The numerical results for the three $z$-selections $z>0.1, z>0.2$ and $0.1<z<0.2$ are available on HepData~\cite{hepdata}.

The TSAs are determined in each of the four $Q^2$-ranges as functions of the variables $x$, $z$ or $p_T$, with the following bin limits:
%
\begin{enumerate}[$x$)))]
 \setlength\itemsep{0.0em}
  \item[$x$:] 0.003, 0.008, 0.014, 0.022, 0.035, 0.055, 0.1, 0.145, 0.215, 0.3, 0.55, 0.9
  \item[$z$:] 0.10,  0.20,  0.30,  0.40,  0.60, 1.0
  \item[$p_T$:] 0.10,  0.30,  0.50,  0.75,  1.0,  7.0 (in units of~\gvc).
\end{enumerate}

The resulting TSAs are carefully studied for possible systematic biases.
The largest systematic uncertainty is due to possible residual acceptance variations within the data taking sub-periods. They are quantified by evaluating various types of false asymmetries. The differences between physical and false asymmetries are used to quantify the overall systematic point-to-point uncertainties, which are evaluated to be 0.5 times the statistical uncertainties.
An additional normalisation uncertainty of $3\%$ originating from the uncertainties of target polarization and dilution factor is not included in the error bands that represent the systematic uncertainties shown in the figures. An additional $5\%$ scale uncertainty has to be added in quadrature for the lepton-polarisation-dependent asymmetries.
More details on analysis and systematic studies can be found in Refs.~\cite{Adolph:2012sp, Adolph:2012sn} and in a forthcoming article~\cite{CMPS10TSA}.
\section{Results and Discussion}
\label{sec:results}
\begin{figure}[tbp]
\centering
\includegraphics[width=1.0\textwidth]{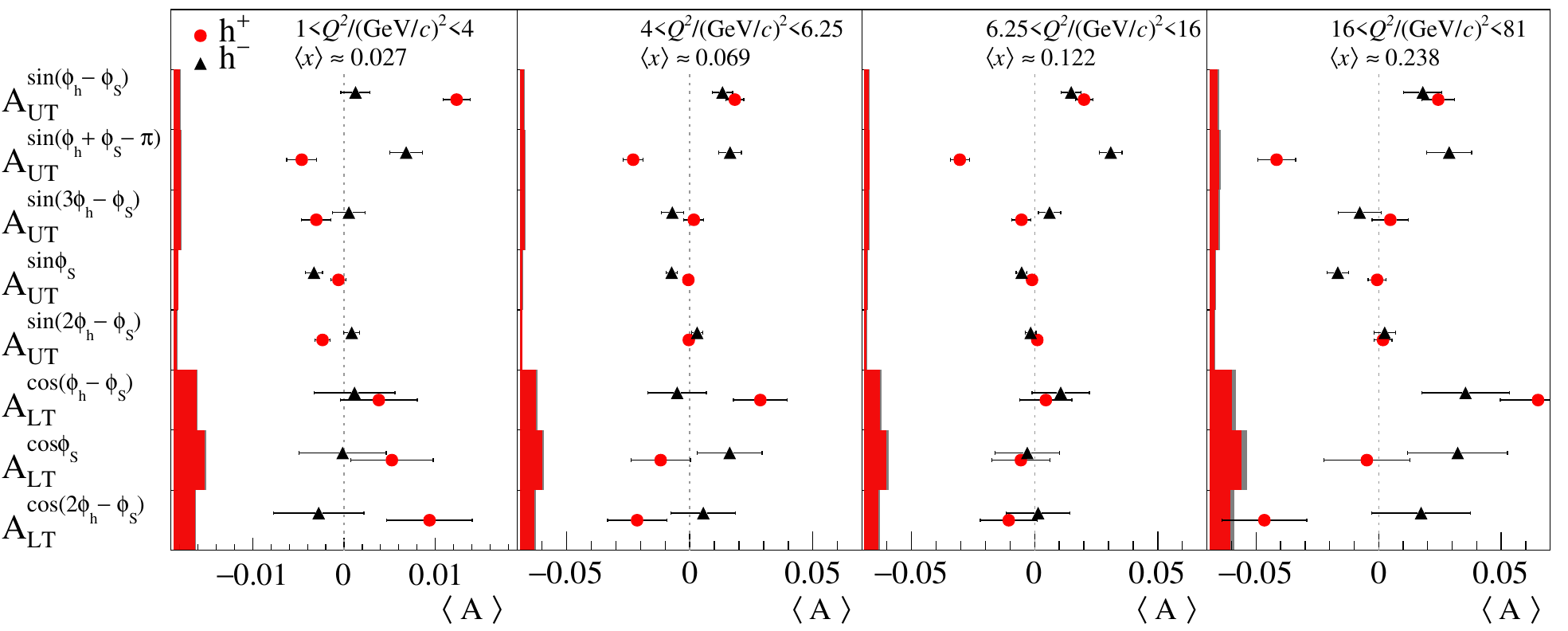}
\caption{Mean TSAs in the four DY $Q^2$-ranges. Error bars represent statistical uncertainties. Systematic uncertainties are shown as error bands next to the vertical axes. For each $Q^2$-range also the average $x$-values are given.}
\label{fig:A8in1}
\end{figure}

The eight TSAs that are extracted from COMPASS SIDIS data in this analysis are shown in Fig.~\ref{fig:A8in1} in the four above defined $Q^2$-ranges, after averaging over all other kinematic dependences. In particular, the Sivers TSA is determined with good statistical accuracy in all four $Q^2$-ranges. For positive hadrons its amplitude is clearly positive in all four $Q^2$-ranges, whereas for negative hadrons it is compatible with zero in the lowest $Q^2$-range and becomes significantly positive in the other three. The other seven TSAs will be discussed in detail in the forthcoming COMPASS article~\cite{CMPS10TSA}, while here they are shown for completeness. The full set of information for all eight TSAs including correlation coefficients and mean kinematic values is available on HepData~\cite{hepdata}.

\begin{figure}[tbp]
\centering
\includegraphics[width=0.9\textwidth]{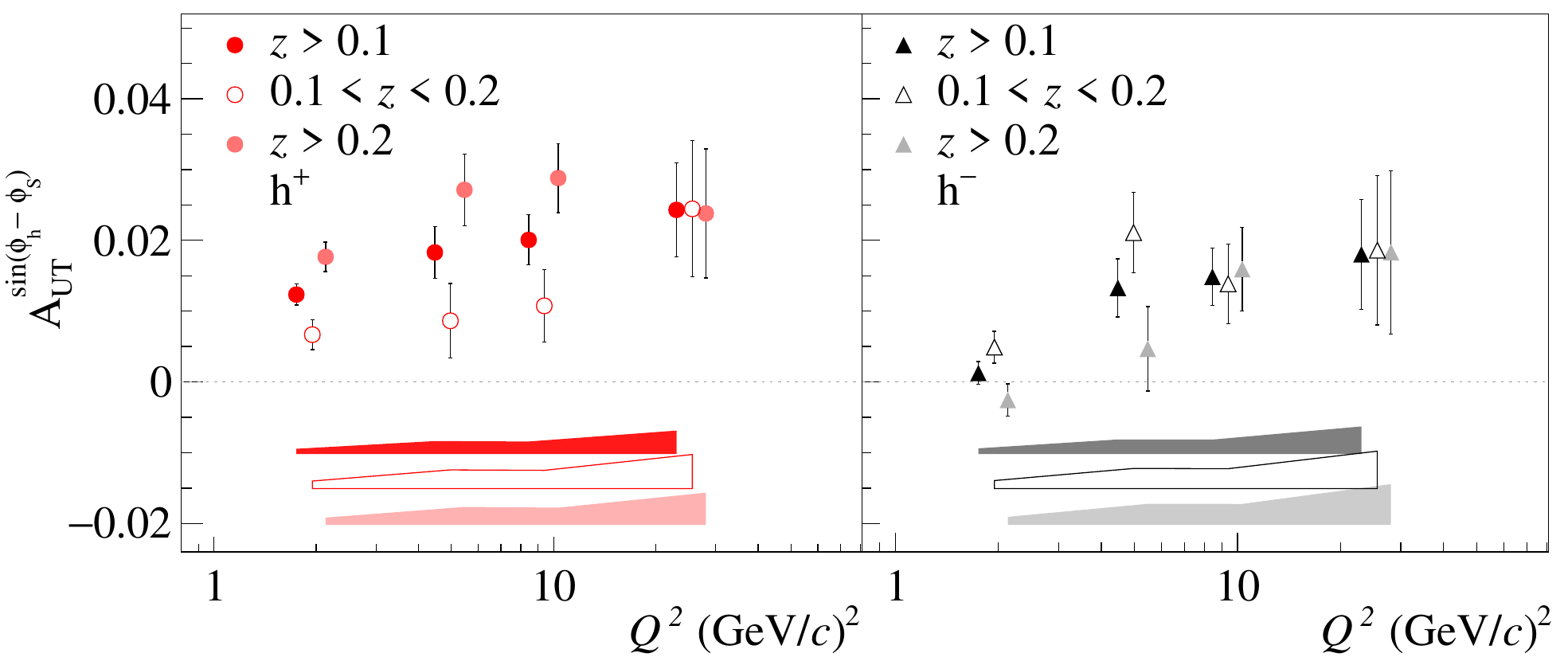}
\caption{The Sivers asymmetry in the four $Q^2$-ranges for positive (left) and negative (right) hadron production for $z>0.1$, $0.1<z<0.2$ and $z>0.2$ ranges. Note that the average $x$-values in these $Q^2$-ranges are different, as can be seen from Fig.~\ref{fig:Q2dist}. The abscissa positions of the points for $z>0.1$ ($z>0.2$) are slightly shifted to the left (right) for better visibility. Error bars represent statistical uncertainties. Systematic uncertainties are shown as bands at the bottom.}
\label{fig:SivZ}
\end{figure}

In Fig.~\ref{fig:SivZ}, the Sivers TSAs for the three $z$-selections are shown after averaging over all other kinematic dependences in each given $Q^2$-range. As it can be seen from this figure, the choice $z>0.1$ maximises the significance of the asymmetry in the highest $Q^2$-range for both positive and negative hadrons and is hence best suited for the determination of the sign of the Sivers TSA in SIDIS.
The increase of the Sivers TSA with $Q^2$ cannot be interpreted as a $Q^2$-dependence as the average $x$-values increase substantially from one $Q^2$-range to the next one, as it can be seen in Fig.~\ref{fig:Q2dist}.

\begin{figure}[tbp]
\centering
\includegraphics[width=0.8\textwidth]{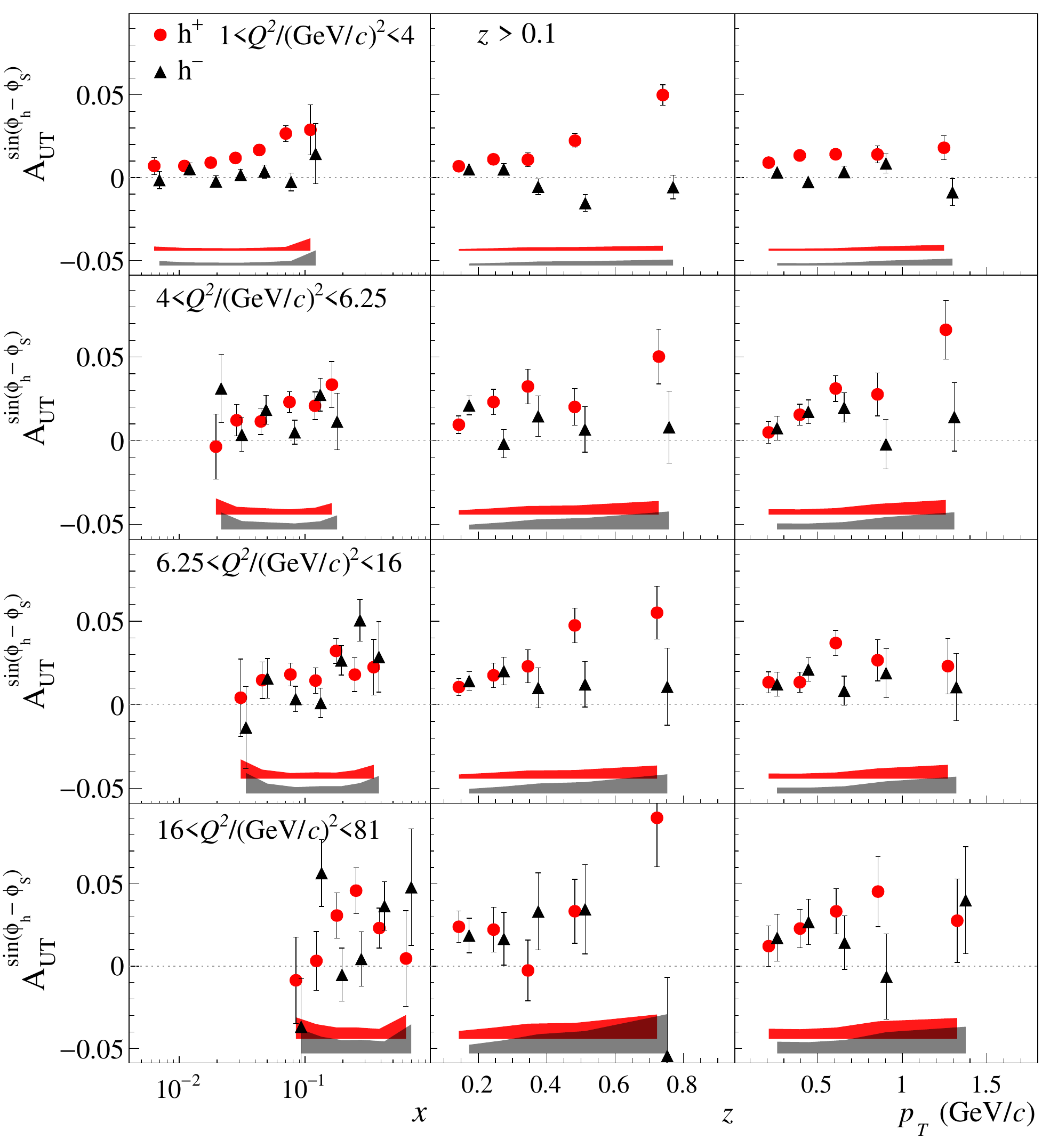}
\caption{Sivers asymmetry for $z>0.1$ in the four $Q^2$-ranges as a function of $x$, $z$ and $p_T$, for positive and negative hadrons. The abscissa positions of the points for negative hadrons are slightly shifted to the right for better visibility. Error bars represent statistical uncertainties. Systematic uncertainties are shown as bands at the bottom.}
\label{fig:Siv}       
\end{figure}

In Fig.~\ref{fig:Siv}, the Sivers TSAs $A_{UT}^{\sin(\phi_h-\phi_S)}$ for positive and negative hadrons are shown as a function of $x$, $z$ and $p_T$ in the four above selected  $Q^2$-ranges.
For positive hadrons, a positive Sivers TSA is observed  in the whole $x$-interval and in all four $Q^2$-ranges (first column). The Sivers asymmetry as a function of $x$ appears to increase up to $x\simeq0.2$ in each of the $Q^2$-ranges, followed by a possible decrease at large $x$. The second and third columns indicate an approximately linear dependence at low $z$ and $p_T$ values. Such a behaviour is supported by the existing phenomenological parametrisations of the Sivers effect~\cite{Anselmino:2005ea, Anselmino:2008sga}.
For negative hadrons, the Sivers TSA is sizeably smaller and less prominent. At intermediate $z$ ($0.3\div0.6$) and low $Q^2$ (first row) it appears to be negative. For larger values of $Q^2$, the Sivers TSA for negative hadrons tends to grow and becomes positive (see also right panel of Fig.~\ref{fig:SivZ}).

\begin{figure}[tbp]
\centering
\includegraphics[width=0.35\textwidth]{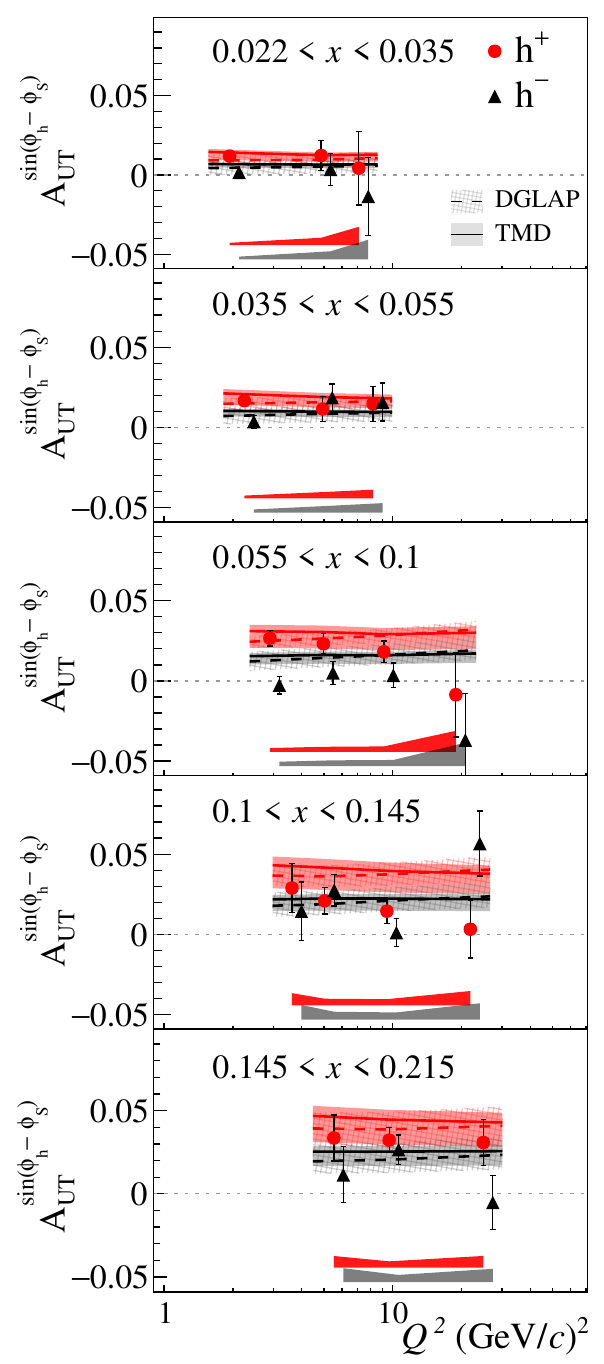}
\caption{The $Q^2$-dependence of the Sivers asymmetry for positive and negative hadrons in five selected bins of $x$. The abscissa positions of the points for negative hadrons are slightly shifted to the right for better visibility. The solid (dashed) curves represent the calculations based on TMD (DGLAP) evolution for the Sivers TSAs~\cite{Anselmino:2012aa, PrivComMEB}. Error bars represent statistical uncertainties. Systematic uncertainties are shown as bands at the bottom.}
\label{fig:SivQ2x}       
\end{figure}

Figure~\ref{fig:SivQ2x} shows the $Q^2$-dependence of the Sivers asymmetry for positive and negative hadrons in five selected bins of x. These are the x-bins to which more than two $Q^2$-ranges contribute. The figure also shows the predictions from collinear (DGLAP) and TMD-evolution, which are based on the best fit~\cite{Anselmino:2012aa} of all published HERMES~\cite{Airapetian:2009ae} and COMPASS~\cite{Alekseev:2008aa, Adolph:2012sp} measurements.
A comparison of the points from the same $x$-bins but different $Q^2$-ranges shows no clear $Q^2$-dependence of the Sivers TSAs within statistical accuracy.
Also, the comparison of fits (not shown in the figure) performed with a linear decreasing function or a constant does not yield a statistically significant conclusion, although there may be a slight preference to the former dependence for positive hadrons. For negative hadrons no clear trend is observed.

In contrast to the DGLAP evolution framework, the present TMD evolution schemes predict a strong $Q^2$-dependence both for polarised and unpolarised TMD PDFs at a given $x$ in fixed-target kinematics. Still, due to partial cancellation of evolution effects in numerator and denominator of the asymmetry, the Sivers TSAs themselves may exhibit only a weak $Q^2$-dependence.
Available descriptions of the Sivers TSAs, which are based on parametrisations of the unpolarised and polarised TMDs, are driven mostly by the one-dimensional data at low $x$ and low $Q^2$ from HERMES and COMPASS, so that present phenomenological studies of $Q^2$-evolution are based on fits using the results of two separate experiments. Present models predict for increasing $Q^2$ a slight increase of the Sivers TSAs for DGLAP and a decrease for TMD evolution.
Based on these fits of one-dimensional data, various TMD-evolution models predict different sizes for the DY Sivers TSA in the high mass range, with values between 0.04 to 0.15~\cite{Aybat:2011ta, Anselmino:2012aa, Sun:2013hua, Echevarria:2014xaa}.
Better constraints on $Q^2$-evolution models of TMDs can be expected only from data that are simultaneously differential in $x$ and $Q^2$, as the data presented in this Letter.

\begin{figure}[tbp]
\centering
\includegraphics[width=0.634\textwidth]{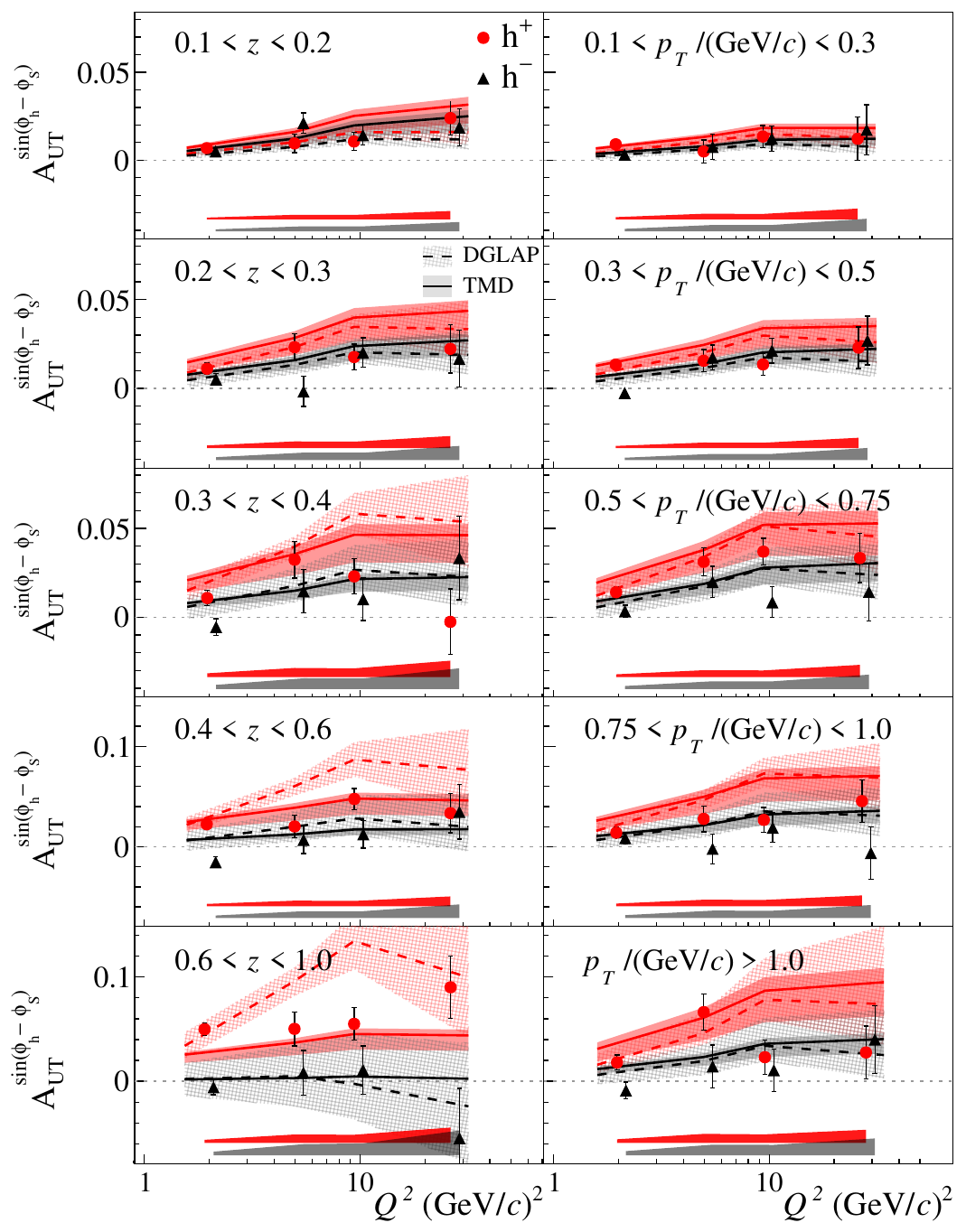}
\caption{The Sivers asymmetry for the four DY $Q^2$-ranges for positive and negative hadrons in bins of $z$ and $p_T$, where the latter is given in units of \gvc. The abscissa positions of the points for negative hadrons are slightly shifted to the right for better visibility. The solid (dashed) curves represent the calculations based on TMD (DGLAP) evolution for the Sivers TSAs~\cite{Anselmino:2012aa, PrivComMEB}. Error bars represent statistical uncertainties. Systematic uncertainties are shown as bands at the bottom.}
\label{fig:SivQ2zpt}       
\end{figure}
In Fig.~\ref{fig:SivQ2zpt}, Sivers TSAs are shown for different $Q^2$-ranges in bins of $z$ and $p_T$.
Note that the average $x$-values in different $Q^2$-ranges are increasing with $Q^2$, as can be seen from Fig.~\ref{fig:Q2dist}. Particularly interesting in Fig.~\ref{fig:SivQ2zpt} is the comparison of the Sivers TSAs for positive and negative hadrons at low $z$ and low $p_T$ (top row). Here, they have small statistical uncertainties and appear to be compatible with one another. Moving towards larger values of $z$ and $p_T$, the two TSAs start to differ.

Fig.~\ref{fig:SivQ2zpt} shows different levels of agreement between our two-dimensional data and the predictions that are based on earlier fits of one-dimensional data~\cite{Anselmino:2005ea, Anselmino:2008sga}.
At low values of $z$ and $p_T$, predictions and data agree within uncertainties.
In particular, there is agreement in the region $0.1<z<0.2$ (top row, left panel), although the corresponding parametrisations were based on a fit to HERMES data in the range $z>0.2$ and $W > \sqrt{10}$~\gvcw~\cite{Airapetian:2009ae} and COMPASS data in the range $z>0.2$ and $W > 5$~\gvcw~\cite{Adolph:2012sp, Adolph:2012sn}. This suggests that at COMPASS kinematics factorisation appears to hold already in the range of low-$z$ and $W > \sqrt{10}$~\gvcw.
At higher values of $z$ and $p_T$, clear discrepancies are observed. In particular, at highest $z$ DGLAP curve for positive hadrons exhibits an apparent artefact at about $Q^2\approx10$~\gvcs. It can be expected that new fits including the two-dimensional Sivers TSAs presented in this Letter will better constrain the models.

\section{Summary and Conclusions}
\label{sec:conclusions}
In this Letter, we present the results of SIDIS measurements of the Sivers TSAs in four different $Q^2$-ranges, chosen to be the same as used in the ongoing analysis of COMPASS DY data.
For the first time, results are given in various two-dimensional $(Q^2,x)$, $(Q^2,z)$, and $(Q^2,p_T)$ representations. For positively charged hadrons, the mean Sivers asymmetry is positive for all four $Q^2$ ranges, while for negatively charged ones it is consistent with zero in the lowest and positive for the other three $Q^2$-ranges.

The range $Q^2>16$~\gvcs~is particularly well suited for the future comparison of COMPASS results on the Sivers effect between SIDIS and DY measurements. It is shown that the SIDIS measurement of the Sivers TSA in this $Q^2$-range yields a positive value with an accuracy that will allow us to test the predicted change of the sign of the Sivers TMD PDF when comparing it to the upcoming results of the analysis of the COMPASS DY measurement in the corresponding range of di-muon mass.

The Sivers TSA measured in the interval $0.1<z<0.2$ agree well with the theoretical predictions that are based on fits on HERMES and COMPASS data, which were obtained for $z>0.2$.
This suggests that at COMPASS kinematics factorisation appears to hold already in the region $z>0.1$.

The observed $Q^2$-dependence of the SIDIS Sivers TSA at given $x$ presently does not allow us to quantitatively distinguish between the predictions for $Q^2$-evolution obtained using TMD and collinear approaches when fitting the existing one-dimensional data.
Future fits using the multi-dimensional data may improve the situation. In this regard, the two-dimensional representations of COMPASS SIDIS TSAs presented in this Letter are the best currently available input from fixed-target experiments.

\section*{Acknowledgements}
\label{sec:aknowledge}
We gratefully acknowledge the support of the CERN management and staff and the skill and effort of the technicians of our collaborating institutes. This work was made possible by the financial support of our funding agencies. Special thanks go to M.~Anselmino, M.~Boglione and A.~Prokudin for providing us with the two-dimensional numerical values of their model predictions and for fruitful discussions.

\bibliography{Sivers_in_DY_bibFile}{}
\bibliographystyle{ws-rv-van}  

\end{document}